\documentstyle[12pt,aaspp4,tighten,psfig,natbib209]{article}
\def\la{\mathrel{\hbox{\rlap{\hbox{\lower4pt\hbox{$\sim$}}}\hbox{$<$}}}}
\def\ga{\mathrel{\hbox{\rlap{\hbox{\lower4pt\hbox{$\sim$}}}\hbox{$>$}}}}

\newcommand{\be}{\begin{eqnarray}}
\newcommand{\ee}{\end{eqnarray}}

\newcommand{\msol}{\ifmmode{{\rm M}_\odot}\else{M$_\odot$}\fi}
\newcommand{\foe}{\ifmmode{10^{51}}\else{$10^{51}$}\fi}

\newcommand{\xni}{\ifmmode{{\rm X}_{\rm Ni}}\else{X$_{\rm Ni}$}\fi}
\def\ang{\hbox{\AA}}
\def\Teff{\ifmmode{T_{\rm eff}}\else{\hbox{$T_{\rm eff}$} }\fi}
\def\Rzero{\ifmmode{R_0}\else{\hbox{$R_0$} }\fi}

\def\SP2{{\tt IBM SP2}}

\def\PC2{{\tt PC$^2$}}

\def\logg{\log(g)}

\def\inu{\ifmmode{I_{\nu}}\else{\hbox{$I_{\nu}$} }\fi}
\def\snu{\ifmmode{S_{\nu}}\else{\hbox{$S_{\nu}$} }\fi}
\def\jnu{\ifmmode{J_{\nu}}\else{\hbox{$J_{\nu}$} }\fi}

\def\etal{et al.}
\def\fep{\ifmmode{{\rm Fe II}}\else\hbox{Fe~II }\fi}

  at 12.0 true pt %scaled \magstep1

\def\etal{{et al}}

\def\phoenix{{\tt PHOENIX}}

\def\etal{{et al}}

\def\phoenix{{\tt PHOENIX}}

\def\b{\beta}

\def\rout{\ifmmode{r_{\rm out}}\else\hbox{$r_{\rm out}$}\fi}
\def\tmax{\ifmmode{\tau_{\rm max}}\else\hbox{$\tau_{\rm max}$}\fi}
\def\tstd{\ifmmode{\tau_{\rm std}}\else\hbox{$\tau_{\rm std}$}\fi}
\def\vmax{\ifmmode{v_{\rm max}}\else\hbox{$v_{\rm max}$}\fi}
\def\muE{\ifmmode{\mu_{\rm E}}\else\hbox{$\mu_{\rm E}$}\fi} 
\def\pE{\ifmmode{p_{\rm E}}\else\hbox{$p_{\rm E}$}\fi} 
\def\bmax{\ifmmode{\b_{\rm max}}\else\hbox{$\b_{\rm max}$}\fi}
\def\kms{\hbox{$\,$km$\,$s$^{-1}$}}

\def\ang{\hbox{\AA}}

\def\Teff{\hbox{$\,T_{\rm eff}$} }

\def\alog#1{\times 10^{#1}}

\def\rout{\hbox{$r_{\rm out}$} }
\def\pgas{\hbox{$P_{\rm gas}$} }
\def\chistd{\ifmmode{\chi_{\rm std}}\else\hbox{$\chi_{\rm std}$}\fi}

\def\k{\,{\rm K}}
\def\msol{$M_\odot$}
\def\foe{10^{51}}

\def\xni{{\rm X}_{\rm Ni}}

\def\lstar{\ifmmode{\Lambda^*}\else\hbox{$\Lambda^*$}\fi} 
\def\Rop{\ifmmode{[R_{ij}]}\else\hbox{$[R_{ij}]$}\fi}

\def\Rji{\ifmmode{[R_{ji}]}\else\hbox{$[R_{ji}]$}\fi}
\def\Rstar{\ifmmode{[R_{ij}^*]}\else\hbox{$[R_{ij}^*]$}\fi}

\def\Rjistar{\ifmmode{[R_{ji}^*]}\else\hbox{$[R_{ji}^*]$}\fi}
\def\DRji{\ifmmode{[\Delta R_{ji}]}\else\hbox{$[\Delta R_{ji}]$}\fi}
\def\DRij{\ifmmode{[\Delta R_{ij}]}\else\hbox{$[\Delta R_{ij}]$}\fi}

\def\ns{\ifmmode{N_{\rm s}}          % Anzahl der tau-punkte
        \else\hbox{$N_{\rm s}$}\fi}

\def\mat#1{{\bf #1}}     % Macro fr Matrizen
\def\vek#1{{#1}}         % Macro fr Vektoren

\newcount\eqcount
\def
  \nummer{
    \global\advance\eqcount by 1
    (\the\eqcount)
  }

\def
  \numadv{
    \global\advance\eqcount by 1
  }

\def
   \numout#1{
     (\the\eqcount #1)
  }

\def\ivek#1#2{\ifmmode{\vek{I}^{#1}_{#2}}
        \else\hbox{$\vek{I}^{#1}_{#2}$}\fi}

\def\tmat#1#2{\ifmmode{\mat{t}^{#1}_{#2}}
        \else\hbox{$\mat{t}^{#1}_{#2}$}\fi}
\def\rmat#1#2{\ifmmode{\mat{r}^{#1}_{#2}}
        \else\hbox{$\mat{r}^{#1}_{#2}$}\fi}
\def\bvek#1#2{\ifmmode{\beta^{#1}_{#2}}
        \else\hbox{$\beta^{#1}_{#2}$}\fi}

\def\lp{\ifmmode{\lambda^+_\tau}           % lambda +
        \else\hbox{$\lambda^+_\tau$}\fi}
\def\lm{\ifmmode\lambda^-_\tau             % lambda -
        \else\hbox{$\lambda^-_\tau$}\fi}

\begin{document}
\bibliographystyle{apj}

\title{NLTE effects of Ti~I in M dwarfs and giants}

\author{Peter H. Hauschildt}
\affil{Dept.\ of Physics and Astronomy, 
University of Georgia, Athens, GA 30602-2451\\
Email: {\tt yeti@hal.physast.uga.edu}\\
and\\
Dept.\ of Physics and Astronomy,
Arizona State University, Tempe, AZ 85287-1504}

\author{France Allard \& David R. Alexander}
\affil{Dept.\ of Physics, Wichita State University,
Wichita, KS 67260-0032\\
E-Mail: \tt allard@eureka.physics.twsu.edu \& dra@twsuvm.uc.twsu.edu}

\author{E. Baron}
\affil{Dept. of Physics and Astronomy, University of Oklahoma
440 W. Brooks, Rm 131, Norman, OK 73019-0225\\
E-Mail: \tt baron@phyast.nhn.ou.edu}

\begin{abstract}

We present detailed NLTE Ti~I calculations in model atmospheres of
cool dwarf and giant stars.  A fully self-consistent NLTE treatment 
for a Ti~I model atom
with 395 levels and 5279 primary bound-bound transitions is included,
and we discuss the implication of departures from LTE in this atom for
the strengths of Ti~I lines and TiO molecular bands in cool star spectra.
We show that in the atmospheric parameter range investigated, LTE is a
poor approximation to Ti~I line formation, as expected from the low
collisional rates in cool stars.  The secondary effects of Ti~I
overionization on the TiO number density and the TiO molecular opacities, however,
are found to be negligible in the {\em molecular} line forming region for
the relatively small parameter range studied in this paper. 
\end{abstract}

\section{Introduction}

%Johnson, H.R. 1994, "Effects of Non-Local Thermodynamic Equilibrium 
%(NLTE) on Molecular Opacities" in Molecules in the Stellar Environment, 
%ed. U. G. Jorgensen (Proc. IAU Colloq. 146), Lecture Notes in  Physics
%438, Springer-Verlag, Berlin, pp. 234-249.

The opacity in the outer layers of the atmospheres of M stars are
dominated by a small number of very strong molecular compounds (H$_2$O,
TiO, H$_2$, CO, VO). Most of the hydrogen is locked in molecular H$_2$,
most of the carbon in CO; and H$_2$O, TiO, and VO opacities define
a pseudo-continuum covering the entire flux distribution of these
stars. The optical ``continuum'' is due to TiO vibrational bands
which are often used as temperature indicators for these stars. The
indicators may be the depth of the bands relative to the peaks between
them; or the depth of the VO bands; or the strength of the atomic
lines relative to the local ``continuum''; or even the strength of
the infrared water bands; all of these depend on the strength of the
TiO bands and the amount of flux-redistribution to longer wavelengths
caused by the blocking of the flux in the optical and near-IR by
TiO. Departures from LTE of the Ti~I atom, and thus indirect changes
in the concentration of the important TiO molecule, could therefore
have severe and measurable consequences on the atmospheric structure
and spectra of cool stars \cite[]{nlte-john,gjreview}.  Departures from
LTE in the photospheres of cool stars have been investigated in detail
in the line transfer and dissociative equilibrium of H$_2$ in red giant
atmospheres \cite[]{lambert68,vernazza81,anderson89}.  NLTE effects of
CO have been investigated in the Sun~\cite[]{thompson73, A&W89} and in
red giant atmospheres~\cite[]{carbon76,A&W91,WAJS94}.  These studies led
to the conclusion that even after the passage of pulsationally driven
shocks the recombination of H$_2$ and CO must proceed rapidly, leading to
number densities very close to LTE values.  \cite{lipav} and \cite{licarl}
showed that the equivalent widths of the Li~I resonance lines are affected
by NLTE effects and have derived NLTE curves of growth for a range of
model parameters. \cite{vaa96} have performed NLTE calculations for Na~I
using a method similar to that of \cite{licarl}.  NLTE effects on the
excitation potential and spectral lines of the much larger Ti atom have
been analyzed in late-type giants by~\cite{ruland80,brown83}.

Due to their very low electron temperatures, the electron density is
extremely low in M stars; the absolute electron densities are even lower
than found in low density atmospheres, such as those of novae and SNe.
Collisions with particles other than electrons, e.g., H$_2$ or helium, are
not as effective as electron collisions in restoring LTE, both because of
their smaller cross-sections and their much smaller thermal velocities.
Therefore, collisional rates which tend to restore LTE, could be very
small in cool stars. This in turn could significantly increase the
importance of NLTE effects in M stars when compared to, e.g., solar type
stars with much higher electron densities and temperatures.

The NLTE treatment of molecules such as H$_2$O and TiO which have
several million transitions is a formidable problem which requires
an efficient method for the numerical solution of the multi-level
NLTE radiative transfer problem.  Classical techniques, such as
the complete linearization or the Equivalent Two Level Atom method,
are computationally prohibitive for large model atoms and molecules.
Currently, the operator splitting or approximate $\Lambda$-operator
iteration (ALI) method~\cite[e.g.,][]{cannon,ryb72, ryb-kalk84,scharmer}
seems to be the most effective way of treating complex NLTE radiative
transfer and rate equation problems.  Variants of the ALI method have been
developed to handle complex model atoms, e.g., Anderson's multi-group
scheme~\cite[]{and87,anderson89} or extensions of the opacity distribution
function method \cite[]{HubLan95}.  However, these methods have
problems if line overlaps are complex or if the line opacity changes
rapidly with optical depth, a situation which occurs in cool stellar
atmospheres. The ALI rate operator formalism, on the other hand, has been
used successfully to treat very large model atoms such as Fe directly
and efficiently~\cite[cf.][]{fe2pap,fe2nova,fe2sn}.

In this paper we discuss NLTE effects of
Ti~I in fully self-consistent models for a few representative M dwarf
and M giant model atmospheres and spectra.

\section{Methods and Models}
% details are in paper I, just summarize most relevant facts

In order to investigate the importance of Ti~I NLTE effects on the
formation of cool star spectra, a full NLTE model calculation is
required. This means that the multi-level NLTE rate equations must be
solved self-consistently and simultaneously with the radiative transfer
and energy equations, including the effects of line blanketing and
of the molecular equation of state. For the purpose of this analysis
we use our multi-purpose stellar atmosphere code \phoenix\ originally
developed for the modeling of the expanding atmospheres of novae and
supernovae, and adapted to conditions found in cool stars by Allard and
Hauschildt (1995\nocite{mdpap}, hereafter AH95).  \phoenix\ \cite[version
8.0,][]{fe2nova,parapap} uses a spherical radiative transfer for giant
models $(\log(g)<3.5)$, and an equation of state (EOS) including more than 600
molecules. In all  models TiO$^+$ (and ZrO$^+$) species are included in
the equations of state according to partition functions published by
Gurvich \& Glushko (1982)\nocite{TiOplus} for a consistent ionization
equilibrium of Ti.  For the strongest ca.\ $4\alog{6}$ atomic \& molecular
lines, we use detailed depth-dependent Voigt profiles with improved
damping constant computation \cite[]{andydipl,vb10pap}, and Gaussian
profiles for an additional $10^{7}$ much weaker lines. In addition,
we include ca.\ 2000 photo-ionization cross sections for atoms and
ions (Verner \& Yakovlev 1995)\nocite{verner95}.  These and related
improvements, as well as the resulting LTE "NextGen" model grids, are
described in detail by \cite{andydipl,vb10pap} and Allard \& Hauschildt
(in preparation).

Both the NLTE and LTE lines for atoms and molecules are treated with a
direct line-by-line method.  We do {\em not} use pre-computed opacity
sampling tables, and thereby allow for computation of pressure-dependent
Voigt profiles, changes in the isotope ratios, and avoid possible
interpolation effects.  We dynamically select the relevant lines
from master line lists at the beginning of each iteration and sum
the contribution of every line within a search window to compute the
total line opacity at {\em arbitrary} wavelength points. The latter
is crucial in NLTE calculations in which the wavelength grid is both
irregular and variable (from iteration to iteration due to changes in
the physical conditions). This approach also allows detailed and depth
dependent line profiles to be used during the iterations. To make this
method computationally efficient, we employ modern numerical techniques,
e.g., vectorized and parallel block algorithms with high data locality
\cite[]{parapap}, and we use on high-end workstations or supercomputers
for the model calculations.  In the calculations we present in this paper,
we have set the micro-turbulent velocity $\xi$ to $2\kms$. We include
LTE lines (i.e., lines of species that are not treated in NLTE) if they
are stronger than a threshold $\Gamma\equiv \chi_l/\kappa_c=10^{-4}$,
where $\chi_l$ is the extinction coefficient of the line at the line
center and $\kappa_c$ is the local b-f absorption coefficient.

\phoenix\ is a full multi-level NLTE code, i.e., NLTE effects are
considered self-consistently in the model calculations, including
the temperature corrections. The temperature corrections and the
convection are treated as described in \cite{mdpap}. Hauschildt \& Baron
(1995\nocite{fe2pap}) have extended the numerical method developed by
Hauschildt (1993)\nocite{casspap} for NLTE calculations with a very
detailed model atom of Fe~II. In this section we describe how we apply
this technique to a detailed Ti~I model atom from the list of NLTE
species already available
in the model calculations.

\subsection{NLTE Calculational Method}

The large number of transitions of the Ti~I ion that have to be included in 
realistic models of the Ti~I NLTE line formation require an efficient method 
for the numerical solution of the multi-level NLTE radiative transfer problem. 
As already mentioned, the Ti~I model atom  described here includes more than 
5200 individual NLTE lines plus a large number of weak background 
transitions. Classical techniques, such as the complete linearization or the 
Equivalent Two Level Atom methods, are computationally prohibitive. In 
addition, we are also modeling moving media (e.g., stellar
winds, novae and supernovae), therefore, 
approaches such as Anderson's multi-group scheme or extensions of the opacity 
distribution function method (\cite{HubLan95}) cannot be applied. Again, 
simple approximations such as the Sobolev method, are very inaccurate in 
problems in which lines overlap strongly and make a significant continuum 
contribution (important for weak lines), as is the case for nova (and SN) 
atmospheres \cite[cf.]{fe2nova,fe2sn}. 

 We use, therefore, the multi-level operator splitting method described by
Hauschildt (1993)\nocite{casspap}. This method solves
the non-grey, spherically symmetric, special relativistic equation
of radiative transfer in the co-moving (Lagrangian) frame using the
operator splitting method described in Hauschildt\nocite{s3pap} (1992).
Details of the method are also described 
in \cite{fe2pap}, so we give here only a summary of the method.

Even with highly effective numerical techniques, the treatment of
possibly more than one million NLTE lines poses a significant
computational problem, in particular in terms of memory usage. In
addition, most lines are very weak and do not contribute
significantly to the radiative rates. However, together, they can
influence the radiation field from overlapping stronger transitions and
should be included as background opacity. Therefore, we separate the
stronger ``primary'' lines from the weaker ``secondary'' lines by
defining a threshold in $\log(gf)$, which can be arbitrarily changed.
Lines with $gf$-values larger than the threshold are treated in detail,
i.e., they are fully included as individual transitions in the radiative
transfer (assuming complete redistribution) and rate equations. In
addition, we include special wavelength points within the profile of the
strong primary lines.

 The secondary transitions are included as background NLTE opacity
 sources but
are not explicitly included in the rate equations. Their cumulative
effect on the rates is included, since the secondary lines are treated by
line-by-line opacity sampling in the solution of the radiative transfer
equation. Note that the distinction between primary and secondary
transitions is only a matter of convenience and technical feasibility. It
is {\em not} a restriction of our method or the computer code but can be
easily changed by altering the appropriate input files.  As more powerful
computers become available, all transitions can be handled as primary
lines by simply changing the input files accordingly.  We do not pose
additional thresholds such as the energy or the statistical weight of
the lower level of a line. However, we include in the selection process
only observed lines between known levels in order to include only lines
with well known $gf$- values. All predicted lines of Kurucz are included
as secondary lines \cite[see][]{fe2pap} for completeness.  For all
primary lines the radiative rates and the ``approximate rate operators''
\cite[]{casspap} are computed and included in the iteration process.

\subsection{The Ti~I model atom}

To construct the Ti~I model atom we have selected the first 34 terms of
Ti~I.  We include all observed levels that have observed b-b transitions
with $\log{(gf)} > -3.0$ as NLTE levels where $g$ is the statistical
weight of the lower level and $f$ is the oscillator strength of the
transition. This leads to a model atom with 395 levels and 5279 primary
transitions treated in detailed NLTE. That is, we solve the complete b-f
\& b-b radiative transfer and rate equations for all these levels
including all radiative rates of the primary lines. A Grotrian diagram
of this model atom is shown in Fig.~\ref{ti1grot}. In addition, we treat
the opacity and emissivity for the remaining $\approx 0.8$ million weak
secondary b-b transitions in NLTE, if one level of a secondary
transition is included in the model. A detailed description of the
numerical method is presented in Hauschildt \& Baron (1995)\nocite{fe2pap}.

Photo-ionization and collisional rates for Ti~I are not yet available.
Thus, we have taken the results of the Hartree Slater central field
calculations of Reilman \& Manson (1979)\nocite{reilman79} to scale
the ground state photo-ionization rate and have then used a hydrogenic
approximation for the energy variation of the cross-section. Although
they are only very rough approximations, the exact values of the b-f
cross-sections are not important for the opacities themselves which
are dominated by known b-b transitions of Ti~I and other species.
They do, however, have an influence on the actual b-f rates but this
remains unimportant for the computational method used in this work.

While collisional rates are important in hotter stellar atmospheres
with high electron densities, they remain nearly negligible when
compared to the radiative rates for the low electron densities found
in cool stars. We have approximated bound-free collisional rates using
the semi-empirical formula of Drawin (1961)\nocite{drawin61}. The
bound-bound collisional rates are approximated by the semi-empirical
formula of Allen (1973)\nocite{allen_aq}, while the Van~Regemorter's
formula (1962\nocite{vr62}) was used for permitted transitions.

A more accurate treatment of this model atom requires the availability
of more accurate collisional and photo-ionization rates for Ti~I. In the
present calculations we have neglected collisions with particles other
than electrons because the cross-sections are basically unknown.
Additional collisional processes would tend to restore LTE, therefore,
the NLTE effects that we obtain in our calculations should be maximized.

\section{Results}

We have computed a small number of both giant and dwarf models to
investigate the effects of Ti NLTE on the structure and the spectra of
cool stars. The giant models are computed using the spherically symmetric,
static mode of \phoenix, while the dwarf models use the plane-parallel
mode. For the three giants we use a gravity of $\log(g)=1.0$, solar
abundances and the following $\Teff=3200\k$, $3600\k$, and $4000\k$.
All giant models have radial extensions of less than 10\%. In addition,
we computed two NLTE dwarf models with $\logg=5.0$, solar abundances
and $\Teff=2700\k$ and $4000\k$ for comparison. All models include the
Ti~I NLTE treatment as discussed above as well as the standard \phoenix\
equation of state (NLTE mode) and additional LTE background lines (about
10--15 million atomic and molecular lines). The NLTE effects are included
in both the temperature iterations (so that the structure of the models
includes NLTE effects) and all radiative transfer calculations. With the
exception of the hottest models, the majority of the line opacity comes
from the LTE TiO and water wapor lines in most layers of the models,
however, the Ti~I lines are an important source of opacity in some of
the outer laters of the models, in particular in the hotter models.
For Ti~I\ primary lines
we use 5 to 11 wavelength points within their profiles.
This procedure typically leads to about 150,000 wavelength 
points for both the model iteration and the synthetic spectrum calculations.

\subsection{Departure coefficients for Ti~I}

In Figs.~\ref{fig-bi-32-giant}---\ref{fig-bi-40-giant} we show
the departure coefficients of Ti~I for the three giant models.
Figures~\ref{fig-bi-27-dwarf} and \ref{fig-bi-40-dwarf} show the
corresponding results for the two dwarf models. The optical
depth scale, $\tstd$, is measured in the b-f and f-f continuum
at $1.2\mu m$. Note that the atmospheres are {\em extremely} non-grey,
some Ti~I lines form at $\tstd \approx 10^{-6}$. Therefore, we show
in the following graphs a large dynamical range of standard optical
depth.

The departure coefficients
in the dwarf models are significantly smaller than in the giant models,
by several orders of magnitude.  This is mostly an ionization effect,
in the dwarf models Ti~I and Ti molecules (mainly TiO) have higher
concentrations relative to Ti~II than in the giant models (see below). The
electron temperatures in the outer regions of the dwarfs and giants are
comparable. Therefore, the departure coefficients $b_i$ are distinctly
different since their definition explicitly includes the concentrations
of both electrons and Ti~II \cite[cf.][p.~219]{mihalas78}.

 For the giant models, the $\Teff=3600\k$ model has the smallest spread
of the departure coefficients compared to the $3200\k$ and $4000\k$
models, indicating the smallest ``internal'' NLTE effects for this model.
The range of the $b_i$'s in all giant models is about a factor of 10 at
$\tstd=10^{-6}$, but the line forming regions are farther inward,
roughly between $\tstd=10^{-4}$ and $\tstd=1$. For most of the Ti~I
levels, the departure coefficients are less than unity, however at lower
effective temperatures more levels show $b_i>1$ in the outer part of the
atmosphere. Typically, $b_i<1$ can be associated with an overionization
of Ti~I relative to the LTE state. However, this is valid {\em only} if
the electron densities of the NLTE and LTE cases are very similar,
otherwise the behavior is more complicated \cite[]{fe2nova}. In the models
presented here, the electron densities of the LTE and NLTE models are
basically identical because we have intentionally treated the important
electron donors (mostly alkali and earth-alkali metals) in LTE.

 The spread of the departure coefficients is much larger in the dwarf
models, in particular for the cooler model. However, the large electron
pressures in the dwarfs will compensate for the effect of the departures
from LTE and result in smaller changes in the Ti~I line profiles, as
we discuss below.

\subsection{NLTE effects on balance of Ti~I, II and Ti-molecules}

 To analyze the effects of Ti~I NLTE on the formation of
the important TiO molecule, the major opacity source in the
optical spectra of cool stars, we plot the relative
concentration $P_i/\pgas$ of a variety of Ti ions and molecules
for both the LTE and the NLTE cases in 
Figs.~\ref{fig32-gianta}---\ref{fig3d}. 
The LTE figures were constructed by using the 
NLTE structure of the model atmosphere but setting all $b_i=1$.
The LTE and NLTE structures are very similar out to $\tstd\approx 10^{-7}$,
in the very outer regions the NLTE models are up to 300K cooler than
their LTE counterparts due to increased line cooling. Therefore, we
have used the NLTE structures exclusively in these figures in order
to avoid the effects of the cooler outer layers on the plots.

 The main differences between the giant and the dwarf models 
are caused by higher pressures in the dwarf models compared
to the giants. This results in Ti~II being more important 
in the giants than in the dwarfs (compare Figs.~\ref{fig40-giantb}
and \ref{fig3d}). The higher pressures in dwarfs 
also cause much higher concentrations of Ti-molecules in
dwarfs than in giants of comparable effective temperatures.

 Comparison of the LTE and NLTE results for the 
same effective temperatures shows that for the giants
the effects on the concentration of Ti species is relatively
small for $\Teff=3200\k$. Only for $\tstd \le 10^{-4}$ do NLTE
effects on the EOS become noticeable. The main effect is that
the concentration of Ti~I is smaller, whereas the concentration
of Ti~II and TiO$^+$ are slightly higher than in the LTE model.
The concentration of TiO is hardly changed by NLTE effects
in this particular model, only in the outermost optically
very thin regions does the TiO concentration drop, with a 
maximum change of about a factor of 2. 

 The results are similar for the $\Teff=3600\k$ giant model.
Here, NLTE effects prevent recombination of Ti~II
to Ti~I in the very outer parts of the atmosphere. However,
the optical depths there are very small and the effects
on the spectrum due to the EOS NLTE effects are quite small. 
As in the $\Teff=3200\k$ model, the TiO concentration is reduced but the
TiO$^+$ concentration
is increased in the regions where NLTE effects are important 
for the EOS. The $\Teff=4000\k$ giant model reacts 
similarly to the $3600\k$ model, but at this effective temperature
Ti~II is the dominant Ti-species even throughout the LTE 
atmosphere.

 In the dwarf models, the situation is very different. 
The higher pressures favor molecules over atoms and ions,
which impacts the effect of Ti~I NLTE on the structure.
The NLTE $\Teff=2700\k$ model shows increases in the 
concentrations of TiO$^+$ and Ti~II for $\tstd < 10^{-2}$ but
virtually no change in the concentrations of the 
dominant TiO and TiO$_2$ molecules. Note the TiO$^+$
increases more than the Ti~II concentration, indicating that
the additional Ti~II initially created by NLTE overionization
of Ti~I is converted into TiO$^+$.

 The $\Teff=4000\k$ dwarf model shows a somewhat different
behavior. In this model, the NLTE effects increase the 
concentration of Ti~II below $\tstd \approx 10^{-5}$ 
significantly so that Ti~II is the dominant Ti-species below
$\tstd \approx 10^{-6}$, whereas in the LTE model TiO 
becomes the dominant species below $\tstd \approx 10^{-7}$.
NLTE effects reduce the overall concentration of Ti-molecules,
but increase the relative importance of TiO$^+$.

 All changes in Ti-molecules occur at small optical
depths in the outer regions of the model atmospheres.
Considering the fact that the molecular lines form
deeper in the atmosphere than the atomic lines, 
this indicates that the effects of atomic Ti NLTE on the
lines of Ti-molecules will be very small.

\subsection{NLTE effects on Ti~I line profiles in M Star Spectra}

% NLTE effects on the formation of Ti~I lines

 The influence of NLTE effects on the formation and emergent profiles of
the Ti~I lines can be estimated from the run of the ratio of the line
source function $S_L$ to the local Planck-function versus the optical
depth. This is shown in a series of overview plots in 
Figs.~\ref{src-a}---\ref{src-dwarf-b} for all models.  
The large number of Ti~I lines in
NLTE masks individual transitions, but the figures show the wide range
that $S_L/B$ spans for the different transitions. In general, the lines
become optically thin for a wide range of optical depths, between $\tstd
\approx 10^{-5}$ and unity. The electron temperatures and thus the
Planck-function drop rapidly with smaller optical depths, creating
large $S_L/B$ ratios in the outer atmosphere. This indicates that
$S_L$ has decoupled completely from the thermal pool above the line
forming region of each line, as expected.

 The $S_L/B$ ratio is overall closest to unity, its LTE value, for the
$\Teff=3600\k$ giant model. Therefore, this model should show the
smallest changes in line profiles of the three giant models. However,
even in this model many lines have $S_L/B$ ratios very different from
unity, therefore a number of Ti~I lines will be different from the LTE case.
For many transitions, the ratio $S_L/B$ is less than unity in the line
forming region, indicating that the absorption lines will show deeper
cores than in the LTE case.  Note that this effect combines with the
effects of the Ti-NLTE on the EOS, which for the models discussed here
generally {\em reduces} the concentration of Ti~I, in part
canceling the effects due to NLTE scattering.

% to demonstrate internal NLTE
% some lines get weaker (UV), some get stronger (scattering?)

% The effects of Ti~I NLTE on the synthetic spectra

 We demonstrate the net effect of NLTE on the formation of Ti~I lines in
giants in Figs.~\ref{spec3a} to \ref{spec3c}.  As expected, NLTE
effects are smallest for the model with $\Teff=3600$.  
In the optical spectral
region, the changes caused by the Ti~I NLTE line formation are very
small and would be hard to observe due to the enormous crowding of lines
in this spectral region.
NLTE effects for the same Ti~I lines are smaller for
dwarfs at similar effective temperature, cf.\ Figs.~\ref{spec4b}
and \ref{spec4a}.

In the cooler models, the Ti~I lines form deeper in the atmosphere in a
region in which the radiation field is nearly Planckian and thus NLTE
effects are very small. This is due to the enormous background opacity
of TiO and other molecules. In the outer atmosphere of the cooler models
the concentration of Ti~I is much smaller than that of TiO, thus the
effect of the large departures from LTE that we find in these regions on
the Ti~I line profiles is very small.

 In the hotter dwarf models, NLTE effects, in particular on the near-IR lines,
are larger. The giant models, however, show an opposite behavior
(the NLTE effects on the Ti~I lines are larger for the cooler models).
In these models the line forming region of the Ti~I
lines is inside the region in which departures from LTE are
significant. In addition, the TiO opacity is relatively smaller than in
the cooler models. For the  Ti~I line at $\lambda_{\rm vac} \approx
9641\ang$, NLTE effects make the core of the line deeper than the LTE
model predicts. This is due to line scattering which removes photons
from the line core and re-distributes them into the line wings.  These
effects are present in most Ti~I lines, but there are a few
exceptions in the giant model with $\Teff=4000$ for which some lines
are
{\em weaker} in NLTE than in LTE (cf.\ Fig.~\ref{spec3c}). Abundance
determinations of Ti, or likely all metals, from near-IR or IR lines
should therefore include NLTE effects wherever possible.

\section{Summary and Conclusions}

%Ti I: no effect on TiO for model params considered, however, some
%  model parameters with huge effects may exist
% line can get weaker (UV) and stronger (optical) for Ti I
% nlte produces more Ti II which in turn allows TiO+ to form more and more
% effects on structure small because TiO doesn't get killed
% maybe molecular NLTE is required?
%

 For the model parameters that we have considered so far, effects of
Ti~I NLTE on the TiO bands are very small. This seems to be due to the
fact that the line forming region of TiO (around $\tstd \approx 10^{-2}$)
is inside of the line forming region of the Ti~I lines.
Therefore, the TiO lines form in a region in which the Ti~I atom is
basically in LTE and Ti~I NLTE effects that are important at smaller
optical depths cannot affect TiO lines significantly. The situation
is different for the TiO$^+$ molecule which forms from Ti~II and O~I
(cf.\ Fig.~\ref{fig3b}). This molecule is very sensitive
to Ti~I NLTE effects and its lines would be helpful indicators of
Ti~I NLTE effects.

 In further investigations it will be very interesting to look for NLTE
effects in the TiO line formation itself. This is certainly feasible
with modern numerical techniques once adequate data for TiO are
available. In addition, NLTE effects of carbon, nitrogen, oxygen,
Ti~I--II and, in hotter models, Fe~I--II need to be investigated in
detail for larger grid of models to assess the impact of NLTE on model
structures and spectra over wide regions of the Hertzsprung-Russell
diagram. We are currently preparing such models and will report the
results elsewhere (Hauschildt \etal, in preparation).

\smallskip
\begin{small}
\noindent{\em Acknowledgments:}
It is a pleasure to thank  H.R. Johnson, S. Starrfield, and S. Shore
for stimulating discussions. We also thank the referee for helpful
comments that helped us improving the paper. This work was supported in
part by NASA ATP NAGW 5-3618 and LTSA NAGW 5-3619 grants to UGA, by NASA
LTSA grants NAGW 4510 and NAGW 2628, NASA ATP grant NAG 5-3067 to
Arizona State University, by NSF grant AST-9417242 and an IBM SUR grant
to University of Oklahoma; as well as by NSF grant AST-9217946 to Indiana
University and by NASA LTSA grant NAG 5-3435 and a
NASA EPSCoR grant to Wichita State University.
Some of the calculations presented in this paper were performed at the
San Diego Supercomputer Center (SDSC) and the Cornell Theory Center
(CTC), with support from the National Science Foundation, and at the
NERSC with support from the DoE. We thank all these institutions for a
generous allocation of computer time.
\end{small}

\bibliography{yeti,opacity,novae,ltsa,mdwarf,radtran,opacity-fa}

\clearpage
\section{Figures}

\begin{figure}[t]
\caption[]{\label{ti1grot}Simplified Grotrian diagram of our Ti~I model atom.  
All 395 levels and 5379 primary (i.e., full NLTE) 
lines are shown but the 0.8 million secondary (approximate) NLTE lines have been 
omitted for clarity.}
\end{figure}

\begin{figure}[t]
% directory: /sue2/yeti/stars/Giants
% IDL file: fig-bi-ti-3200-giant.pro
%\psfig{file=fig-bi-ti-3200-giant.ps,width=\hsize,angle=90}
\caption[]{\label{fig-bi-32-giant}
Run of the Ti departure coefficients 
for the $\Teff=3200\,$K, $\log g=1.0$ giant model.}
\end{figure}

\begin{figure}[t]
% directory: /sue2/yeti/stars/Giants
% IDL file: fig-bi-ti-3600-giant.pro
%\psfig{file=fig-bi-ti-3600-giant.ps,width=\hsize,angle=90}
\caption[]{\label{fig-bi-36-giant}
Run of the Ti departure coefficients 
for the $\Teff=3600\,$K, $\log g=1.0$ giant model.}
\end{figure}

\begin{figure}[t]
% directory: /sue2/yeti/stars/Giants
% IDL file: fig-bi-ti-4000-giant.pro
%\psfig{file=fig-bi-ti-4000-giant.ps,width=\hsize,angle=90}
\caption[]{\label{fig-bi-40-giant}
Run of the Ti departure coefficients 
for the $\Teff=4000\,$K, $\log g=1.0$ giant model.}
\end{figure}

\begin{figure}[t]
% directory: /sue2/yeti/stars/ti-nlte
% IDL file: plotbi.pro
%\psfig{file=fig-bi-ti-2700-dwarf.ps,width=\hsize,angle=90}
\caption[]{\label{fig-bi-27-dwarf}
Run of the Ti departure coefficients 
for the $\Teff=2700\,$K, $\log g=5.0$ dwarf model.}
\end{figure}

\begin{figure}[t]
% directory: /sue2/yeti/stars/ti-nlte
% IDL file: plotbi.pro
%\psfig{file=fig-bi-ti-4000-dwarf.ps,width=\hsize,angle=90}
\caption[]{\label{fig-bi-40-dwarf}
Run of the Ti departure coefficients 
for the $\Teff=4000\,$K, $\log g=5.0$ dwarf model.}
\end{figure}

\clearpage

\begin{figure}[t]
% directory: /home/hobbes/yeti/MODELS/Giants-nlte
% IDL file: fig-prs-3200-giant.pro
%\psfig{file=fig-prs-3200-giant.ps,width=\hsize,angle=90}
\caption[]{\label{fig32-gianta}
NLTE effects on the Ti ionization and molecule formation
for the $\Teff=3200\,$K, $\log g=1.0$ giant model. The thin
lines and symbols give the results for the LTE model, whereas the
thick lines and symbols give the results for corresponding NLTE
model.}
\end{figure}

\begin{figure}[t]
% directory: /home/hobbes/yeti/MODELS/Giants-nlte
% IDL file: fig-prs-3600-giant.pro
%\psfig{file=fig-prs-3600-giant.ps,width=\hsize,angle=90}
\caption[]{\label{fig36-giantb}
NLTE effects on the Ti ionization and molecule formation
for the $\Teff=3600\,$K, $\log g=1.0$ giant model. The thin
lines and symbols give the results for the LTE model, whereas the
thick lines and symbols give the results for corresponding NLTE
model.}
\end{figure}

\begin{figure}[t]
% directory: /home/hobbes/yeti/MODELS/Giants-nlte
% IDL file: fig-prs-4000-giant.pro
%\psfig{file=fig-prs-4000-giant.ps,width=\hsize,angle=90}
\caption[]{\label{fig40-giantb}
NLTE effects on the Ti ionization and molecule formation
for the $\Teff=4000\,$K, $\log g=1.0$ giant model. The thin
lines and symbols give the results for the LTE model, whereas the
thick lines and symbols give the results for corresponding NLTE
model.}
\end{figure}

\clearpage

\begin{figure}[t]
% directory: /home/hobbes/yeti/MODELS/ti-nlte
% IDL file: fig-prs-2700-dwarf.pro
%\psfig{file=fig-prs-2700-dwarf.ps,width=\hsize,angle=90}
\caption[]{\label{fig3b}NLTE effects on the Ti ionization and molecule formation
for the $\Teff=2700\,$K, $\log g=5.0$ dwarf model. The thin
lines and symbols give the results for the LTE model, whereas the
thick lines and symbols give the results for corresponding NLTE
model.}
\end{figure}

\begin{figure}[t]
% directory: /home/hobbes/yeti/MODELS/ti-nlte
% IDL file: fig-prs-4000-dwarf.pro
%\psfig{file=fig-prs-4000-dwarf.ps,width=\hsize,angle=90}
\caption[]{\label{fig3d}NLTE effects on the Ti ionization and molecule formation
for the $\Teff=4000\,$K, $\log g=5.0$ dwarf model. The thin
lines and symbols give the results for the LTE model, whereas the
thick lines and symbols give the results for corresponding NLTE
model.}
\end{figure}

\begin{figure}[t]
% directory: /sue2/yeti/stars/Giants/
% idl file: linefig.pro
%\psfig{file=fig-ti-src-3200-giant.ps,width=\hsize,clip=,angle=90}
\caption[]{\label{src-a}Line source functions for all Ti~I NLTE lines
for a giant model with the parameters
$\Teff=3200\,$K, $\log g=1.0$.}
\end{figure}

\begin{figure}[t]
% directory: /sue2/yeti/stars/Giants/
% idl file: linefig.pro
%\psfig{file=fig-ti-src-3600-giant.ps,width=\hsize,clip=,angle=90}
\caption[]{\label{src-b}Line source functions for all Ti~I NLTE lines
for a giant model with the parameters
$\Teff=3600\,$K, $\log g=1.0$.}
\end{figure}

\begin{figure}[t]
% directory: /sue2/yeti/stars/Giants/
% idl file: linefig.pro
%\psfig{file=fig-ti-src-4000-giant.ps,width=\hsize,clip=,angle=90}
\caption[]{\label{src-c}Line source functions for all Ti~I NLTE lines
for a giant model with the parameters
$\Teff=4000\,$K, $\log g=1.0$.}
\end{figure}

\begin{figure}[t]
% directory: /sue2/yeti/stars/ti-nlte/
% idl file: linefig.pro
%\psfig{file=fig-ti-src-2700-dwarf.ps,width=\hsize,clip=,angle=90}
\caption[]{\label{src-dwarf-a}Line source functions for all Ti~I NLTE lines
for a dwarf model with the parameters
$\Teff=2700\,$K, $\log g=5.0$.}
\end{figure}

\begin{figure}[t]
% directory: /sue2/yeti/stars/ti-nlte/
% idl file: linefig.pro
%\psfig{file=fig-ti-src-4000-dwarf.ps,width=\hsize,clip=,angle=90}
\caption[]{\label{src-dwarf-b}Line source functions for all Ti~I NLTE lines
for a dwarf model with the parameters
$\Teff=4000\,$K, $\log g=5.0$.}
\end{figure}

\begin{figure}[t]
% directory: /sue2/yeti/stars/Giants/
% idl file: fig-ti-3200.pro
%\psfig{file=fig-ti-3200.ps,width=\hsize,clip=,angle=90}
\caption[]{\label{spec3a}NLTE effects on the Ti~I line at $\lambda_{\rm
vac} \approx9641\ang$ and around $5020\ang$
for a giant model with the parameters
$\Teff=3200\,$K, $\log g=1.0$. The LTE spectrum (dotted curve)
uses the same model structure as the NLTE spectrum (full curve) but with all
departure coefficients set to unity. The fluxes are in arbitrary units.}
\end{figure}

\begin{figure}[t]
% directory: /sue2/yeti/stars/Giants/
% idl file: fig-ti-3600.pro
%\psfig{file=fig-ti-3600.ps,width=\hsize,clip=,angle=90}
\caption[]{\label{spec3b}NLTE effects on the Ti~I line at $\lambda_{\rm
vac} \approx9641\ang$ and around $5020\ang$
for a giant model with the parameters
$\Teff=3600\,$K, $\log g=1.0$. The LTE spectrum (dotted curve)
uses the same model structure as the NLTE spectrum (full curve) but with all
departure coefficients set to unity. The fluxes are in arbitrary units.}
\end{figure}

\begin{figure}[t]
% directory: /sue2/yeti/stars/Giants/
% idl file: fig-ti-4000.pro
%\psfig{file=fig-ti-4000.ps,width=\hsize,clip=,angle=90}
\caption[]{\label{spec3c}NLTE effects on the Ti~I line at $\lambda_{\rm
vac} \approx9641\ang$ and around $5020\ang$
for a giant model with the parameters
$\Teff=4000\,$K, $\log g=1.0$. The LTE spectrum (dotted curve)
uses the same model structure as the NLTE spectrum (full curve) but with all
departure coefficients set to unity. The fluxes are in arbitrary units.}
\end{figure}

\begin{figure}[t]
% directory: /sue2/yeti/stars/ti-nlte
% idl file: fig-ti-2700-dwarf.pro
%\psfig{file=fig-ti-2700-dwarf.ps,width=\hsize,clip=,angle=90}
\caption[]{\label{spec4b}NLTE effects on the Ti~I line at $\lambda_{\rm
vac} \approx9641\ang$ and around $5020\ang$
for a dwarf model with the parameters
$\Teff=2700\,$K, $\log g=5.0$. The LTE spectrum (dotted curve)
uses the same model structure as the NLTE spectrum (full curve) but with all
departure coefficients set to unity. The fluxes are in arbitrary units.}
\end{figure}

\begin{figure}[t]
% directory: /sue2/yeti/stars/ti-nlte
% idl file: fig-ti-4000-dwarf.pro
%\psfig{file=fig-ti-4000-dwarf.ps,width=\hsize,clip=,angle=90}
\caption[]{\label{spec4a}NLTE effects on the Ti~I line at $\lambda_{\rm
vac} \approx9641\ang$ and around $5020\ang$
for a dwarf model with the parameters
$\Teff=4000\,$K, $\log g=5.0$. The LTE spectrum (dotted curve)
uses the same model structure as the NLTE spectrum (full curve) but with all
departure coefficients set to unity. The fluxes are in arbitrary units.}
\end{figure}

\end{document}